\documentclass[a4paper]{jpconf}
\usepackage{graphicx,cite}

\usepackage{epsfig,multicol,multirow,cite}
\usepackage{amsmath,subfigure,latexsym,amssymb}

\def\backder{\raise1.4ex\hbox{$\leftarrow$\kern-0.75em\raise-1.4ex\hbox{$\partial$}}}
\def\forder{\raise1.4ex\hbox{$\rightarrow$\kern-0.8em\raise-1.4ex\hbox{$\partial$}}}
\newcommand{\backderi}{\mathop{\backder}}
\newcommand{\forderi}{\mathop{\forder}}

\newcommand{\be}{\begin{equation}}
\newcommand{\ee}{\end{equation}}
\newcommand{\nn}{\nonumber}
\newcommand{\bea}{\begin{eqnarray}}
\newcommand{\eea}{\end{eqnarray}} 

\newcommand{\la}{\langle}
\newcommand{\ra}{\rangle}
\newcommand{\uno}{1 \!\! 1}
\newcommand{\Z}{\mathbb{Z}}
\newcommand{\R}{{\kern+.25em\sf{R}\kern-.78em\sf{I} \kern+.78em\kern-.25em}}
\newcommand{\RR}{{\kern+.25em\sf{R}\kern-.6em\sf{I} \kern+.6em\kern-.25em}}
\newcommand{\N}{{\kern+.25em\sf{N}\kern-.78em\sf{I} \kern+.78em\kern-.25em}}
\newcommand{\C}{{\kern+.25em\sf{C}\kern-.50em\sf{I} \kern+.50em\kern-.25em}}

\newcommand{\ri}{{\rm i}}

\makeatletter
\@addtoreset{equation}{section}
\makeatother

\begin{document}
\title{Scalar fields in a non-commutative space}

\author{Wolfgang Bietenholz$^{\rm a}$, Frank Hofheinz$^{\rm b}$, \\
H\'{e}ctor Mej\'{\i}a-D\'{\i}az$^{\rm a}$ and Marco Panero$^{\rm c}$}

\address{$^{\rm a}$ Instituto de Ciencias Nucleares,
Universidad Nacional Aut\'{o}noma de M\'{e}xico \\
\ \ \ A.P.\ 70-543, C.P.\ 04510 M\'{e}xico, Distrito Federal, Mexico \\
$^{\rm b}$ Institute of Radiopharmaceutical Cancer Research \\
\ \ \ Helmholtz-Zentrum Dresden-Rossendorf, Germany \\
$^{\rm c}$ Instituto de F\'{\i}sica Te\'{o}rica UAM/CSIC, 
Universidad Aut\'{o}noma de Madrid \\
\ \ \ Ciudad Universitaria de Cantoblanco, 28049 Madrid, Spain} 

\ead{wolbi@nucleares.unam.mx}

\begin{abstract} 
We discuss the $\lambda \phi^{4}$ model in 2- and 3-dimensional 
non-commutative spaces. The mapping onto a Hermitian matrix model 
enables its non-perturbative investigation by Monte Carlo 
simulations. The numerical results reveal a phase where stripe 
patterns dominate. In $d=3$ we show that in this phase the 
dispersion relation is deformed in the IR regime, in agreement 
with the property of UV/IR mixing. This ``striped phase'' also
occurs in $d=2$. For both dimensions we provide evidence 
that it persists in the simultaneous limit to the continuum 
and to infinite volume (``Double Scaling Limit''). This implies
the spontaneous breaking of translation symmetry.
\end{abstract}

\section{Quantum physics in a non-commutative space}

Since in standard quantum mechanics the operators of space and 
momentum coordinates do not commute, it seems like an obvious
idea to ``quantise further'' by also introducing non-zero
commutators among space coordinates (or among momentum coordinates)
in different directions. Hence it is not surprising that this idea
dates back to the 1940s. Its pre-history involves famous names 
like Heisenberg, Pauli, Peierls and Oppenheimer, and in 1947
the first papers on this subject were published \cite{Sny}.

However, the consequences of non-commutative (NC) geometry in
quantum field theory are extremely involved. Here we consider 
only the simplest case of two NC spatial coordinates,
with a constant non-commutativity ``tensor'' $\Theta$,
\be  \label{NCplane}
[ \hat x_{i} , \hat x_{j} ] = \ri \Theta_{i j} =
\ri \theta \epsilon_{i j} \ ,
\ee
where $\hat x_{i}$ are Hermitian operators, $i,j \in  
\{ 1, 2\}$, and $\theta$ is the non-commutativity parameter.

In the 1980s deep mathematical work was carried out about the formal 
formulation of field theories on such spaces (for a review, see
Ref.\ \cite{Connes}).
Applications were discussed in solid state physics, in particular 
related to the quantum Hall effect, see {\it e.g.}\ Ref.\ 
\cite{Girvin}. Of course, in this context one 
actually deals with the usual geometry, but a magnetic background 
field $B$ can be interpreted as $\theta \propto 1/B$, which leads 
to NC canonical coordinates.

In the period from 1996 to 1998 a boom of interest in NC field
theories set in, which led to about 3000 papers on this subject up 
to now \cite{inspire}. This boom was triggered by the observation
that low energy string theory can be related to NC field theory
\cite{SeiWit}, following the spirit of the re-interpretation of 
the magnetic background field.

Here we are going to address NC field theory as such; no strings
attached. A qualitative
difference from usual field theory is its {\em non-locality;} fields
interact at distinct points over a characteristic distance
$\sim \sqrt{\Vert \Theta \Vert }$. This entails frightening
conceptual problems. On the other hand, from a very optimistic 
point of view, this is just what it takes for a proposal to 
formulate quantum gravity.

In this regard, we mention a simple {\it Gedankenexperiment:}
assume some event to be measured with extremely tiny space-time
uncertainties $\Delta x_{1}, \, \Delta x_{2}, \, \Delta x_{3}, \, 
\Delta t$, on the order of the Planck length $l_{\rm Planck}
\simeq 1.6 \cdot 10^{-35}~{\rm m}$.
This requires a huge energy density, so gravitation should 
be taken into consideration. In the extreme case this could 
yield an event horizon, which is larger
than the Heisenberg uncertainties, so the event is invisible. 
According to Ref.\ \cite{DFR}, avoiding that scenario 
requires
\bea
\Delta x_{1} \Delta x_{2} + \Delta x_{1} \Delta x_{3}
+ \Delta x_{2} \Delta x_{3} & \geq & l_{\rm Planck}^{2} \ , \nn \\
( \Delta x_{1} + \Delta x_{2} + \Delta x_{3}) \Delta t
& \geq & l_{\rm Planck}^{2} \ .
\eea
Such space-time uncertainties are characteristic for an NC space.
Is this a natural framework for the conciliation of quantum
theory and gravity? We should add, however, that much of the
literature on this subject keeps the time commutative. That
deviates from the above consideration, but it may save unitarity
and reflection positivity, and it alleviates the problems
related to causality. \\

The historic motivation, however, was different. People hoped that 
washing out the space-time points in this way\footnote{Some people
denote it as ``pointless geometry''.} 
could remove (or at least weaken) the notorious UV divergences in 
quantum field theory, and avoid (or simplify) renormalisation. This 
turned out to be wrong: renormalisation is not getting easier, but 
much harder due to non-commutativity. First of all, in planar diagrams
of a perturbative expansion, the UV divergences simply persist 
\cite{Filk}. Second, in the non-planar diagrams they tend to ``mix'' 
with IR divergences. This type of singularities does not
occur in the commutative world, and it is very difficult to deal with. 
For a simple intuitive picture, we start again from the Heisenberg 
uncertainty $\Delta x_{i} \sim 1 / \Delta p_{i}$, and we combine it with 
\be
\Delta x_{i} \sim \Theta_{ij}/\Delta x_{j} \sim 
\Theta_{ij} \Delta p_{j} \qquad (i \neq j) \ .
\ee
Therefore an attempt to squeeze $\Delta p_{i} \to 0$
makes $\Delta p_{j}$ diverge. Due to such mixed singularities, 
the renormalisation of multi-loop diagrams is mysterious. 
Hence it is highly motivated to adopt
a fully {\em non-perturbative approach.} \\

In many models, the lattice regularisation enables the non-perturbative 
treatment of quantum field theory. So let us introduce a lattice structure 
also on the NC plane of eq.\ (\ref{NCplane}) (this concept is
reviewed in Ref.\ \cite{Szabo}). This is achieved --- at least in 
a fuzzy form --- if we impose the operator identity
\be
\exp \Big( \ri \frac{2 \pi}{a} \hat x_{i} \Big) = \hat \uno \ .
\ee
For lattice spacing $a$ we expect periodicity of the (commutative)
momentum components over the Brillouin zone,
\be
\exp \Big( \ri \sum_{i} k_{i} \hat x_{i} \Big) = \exp \Big(
\ri \sum_{i} (k_{i} + \frac{2\pi}{a} ) \hat x_{i} \Big) \ ,
\quad i = 1,2 \ .
\ee 
Multiplication with the inverse factor 
$\exp ( - \ri \sum_{j} k_{j} \hat x_{j} )$ from the right,
and applying the Baker-Campbell-Hausdorff formula, leads to
\be
\hat \uno = \hat \uno \, \exp \Big( \frac{\ri \pi}{a}
\theta (k_{2} - k_{1} ) \Big) \quad \Rightarrow \quad
\frac{\theta}{2a} k_{i} \in \Z \ .
\ee
Therefore, in contrast to the commutative space, the NC lattice
is automatically {\em periodic.} 

If we now assume periodicity over the lattice volume $N \times N$,
we have discrete momenta $k^{(n)} = \frac{2 \pi}{a N} n$, 
$n \in \Z^{2}$, and we arrive at the relation
\be
\theta = \frac{1}{\pi} N a^{2} \ .
\ee
In order to keep $\theta$ finite, in particular for
$\theta = {\rm const.}$, we have to take the limits to the 
continuum, $a \to 0$, and to infinite volume, $Na \to \infty$, 
{\em simultaneously.} This is the {\em Double Scaling Limit,}
\be  \label{DSL}
\left.
\begin{array}{ccc}
a & \to & 0 \\
N & \to & \infty
\end{array}
\right\}
\quad {\rm such~that} \quad Na^{2} = {\rm const.} \ ,
\ee
which leads to a NC plane of infinite extent. Clearly,
this requirement is again related to the property of
UV/IR mixing. Taking these limits differently, one would
usually end up with $\theta =0$ or $\theta = \infty$, which 
are both (different) cases of commutative field theory.
For a consistent study of NC field theory, we have to
follow the instruction (\ref{DSL}). 

\section{The non-commutative $\lambda \phi^{4}$ model}

\subsection{Formulation}

NC field theory can be formulated such that the fields are
functions of the standard (commutative) coordinates $x_{\mu}$,
if all field multiplications are performed by {\em star products}
(or {\em Moyal products}). A prototype reads
\be
\phi (x) \star \psi (x) : = \phi (x) \exp \Big( \frac{\ri}{2}
\backderi \,\! _{\mu} \,
\Theta_{\mu \nu} \forderi \,\! _{\nu} \, \Big) \psi (x)
\ee
(for instance $[x_{\mu},x_{\nu}]_{\star} = \ri \Theta_{\mu \nu}$).
This can be justified with a plane wave decomposition.

For bilinear terms under a space-time integral (without boundary
terms) the star product is equivalent to a simple product, because
of the anti-symmetry of $\Theta$. Hence the action of the
$\lambda \phi^{4}$ model in NC Euclidean space can be written as
\be
S [ \phi ] = \int d^{d}x \, \Big[ \frac{1}{2}
\partial_{\mu} \phi \, \partial_{\mu} \phi + \frac{m^{2}}{2} \phi^{2}
+ \frac{\lambda}{4} \phi \star \phi \star \phi \star \phi \Big] \ .
\ee
This shows that the parameter $\lambda$ does not only determine
the strength of the self-interaction, but also the extent
of NC effects.

The perturbative expansion of this model has been discussed
extensively in the literature. Regarding the 1-loop diagrams,
there is a planar contribution, which takes the standard form 
\cite{Filk}, as we mentioned before. On the other hand, the non-planar
diagrams pick up a phase factor due to the non-commutativity.
For the moment, let us introduce a momentum cutoff $\Lambda$.
Then the 1-loop integrals and their leading divergences in $d=4$ 
take the form \cite{Ram}
\be
{\rm planar:} \quad \int d^{d}k \, \frac{1}{k^{2} + m^{2}} 
\propto \Lambda^{2} \ , \quad 
{\rm non\mbox{-}planar:} \quad \int d^{d}k \, 
\frac{\exp (\ri k_{\mu} \Theta_{\mu \nu} p_{\nu})}{k^{2} + m^{2}}
\propto \frac{1}{1/\Lambda^{2} + p_{\mu} \Theta_{\mu \nu} p_{\nu}} \ .
\  \nn
\ee
We see that a finite $\Theta$ does indeed allow us to
take the limit $\Lambda \to \infty$ in the non-planar part.
However, then a singularity emerges for external momentum
$p \to 0$, which illustrates the UV/IR mixing. Moreover, 
even at finite $p$, the limit $\Theta \to 0$ is not smooth; therefore
the meaning of a truncated expansion in small $\Vert \Theta \Vert$ 
is questionable. Finally we also confirm that the opposite
limit $\Theta \to \infty$ is commutative, but different from
$\Theta = 0$.\\

Now we consider $d=3$, so that the scalar field $\phi (\vec x, t)$
lives on a NC plane $(x_{1},x_{2})$, plus a commutative Euclidean 
time $t$. We assume a lattice structure, for the NC plane
as described in Section 1, and for the time in a
regular form. The action on a $N^{2} \times T$ lattice
can be mapped (with identical algebras) onto a matrix model 
with twisted boundary conditions \cite{AMNS},
\be
S [ \bar \phi ] = {\rm Tr} \sum_{t=1}^{T} \Big[ \ \frac{1}{2}
\sum_{i=1}^{2} \Big( \Gamma_{i} \bar \phi (t) \Gamma_{i}^{\dagger} -
\bar \phi(t) \Big)^{2} + \frac{1}{2} \Big( \bar \phi (t+1) - \bar \phi (t)
\Big)^{2} 
+ \frac{m^{2}}{2} \bar \phi^{2}(t) + \frac{\lambda}{4} 
\bar \phi^{4}(t) \ \Big] \ ,
\ee
where $\bar \phi (t)$ are Hermitian $N \times N$ matrices, 
located at \ $t =1 , \dots , T$ (we are using lattice units). 
The kinetic term has the
standard lattice form in time direction, but in the NC plane
it is constructed by means of so-called twist eaters 
$\Gamma_{i}$. They arrange for a shift by one lattice unit,
if they obey the 't Hooft-Weyl algebra
\be  \label{tHW}
\Gamma_{i} \Gamma_{j} = Z_{ji} \Gamma_{j} \Gamma_{i} \ .
\ee
Here the twisted boundary conditions enter, and we choose the
corresponding phase factor as 
\be
Z_{21} = Z_{12}^{*}= \exp \Big( \ri \pi (N+1)/N \Big) \ ,
\ee
where the size $N$ is assumed to be odd. Then we insert a 
unitary solution for $\Gamma_{1}$ and $\Gamma_{2}$, which is
known as clock- and shift-matrix (they are written
down explicitly {\it e.g.}\ in Refs.\ \cite{AMNS,Frank}). The 
crucial property, however, is relation (\ref{tHW}).

\subsection{Phase diagram}

Some years ago, Gubser and Sondhi performed a 1-loop
calculation in the Hartree-Fock approximation \cite{GuSo} and 
conjectured the following properties of the phase diagram of 
the NC $\lambda \phi^{4}$ model in $d=3$ and $4$:

\begin{itemize}

\item At small $\theta$, there is an Ising-type phase transition
between a disordered and a uniform phase at some critical
value $m_{c}^{2} < 0$
(a strongly negative parameter $m^{2}$ can be interpreted as
low temperature).

\item At large $\theta$ and some $m_{c}^{2} < 0$, there is another
phase transition, but now between a disordered and a {\em striped 
phase.}

\end{itemize}

Further considerations were added with Renormalisation Group
techniques \cite{ChenWu}, and with the Cornwall-Jackiw-Tomboulis 
effective action approach \cite{CaZa03}. They are consistent 
with the qualitative picture by Gubser and Sondhi.

\begin{figure}[hbt!]
\vspace*{-5mm}
\center
\includegraphics[angle=0,width=0.8\linewidth]{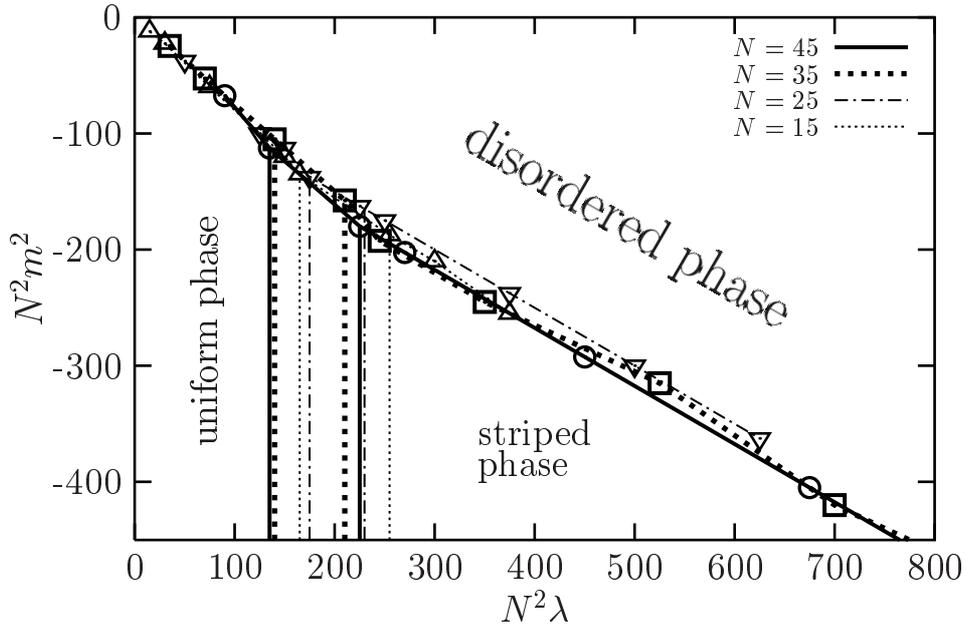}
\caption{The phase diagram of the 3d $\lambda \phi^{4}$ model
on an NC plane but with a commutative Euclidean time coordinate.
The ordered regime --- at strongly negative $N^{2} m^{2}$ --- is 
divided into phases of uniform and of striped order.}
\label{phasedia3d}
\end{figure}

\begin{figure}[hbt!]
\center
\includegraphics[angle=0,width=0.33\linewidth]{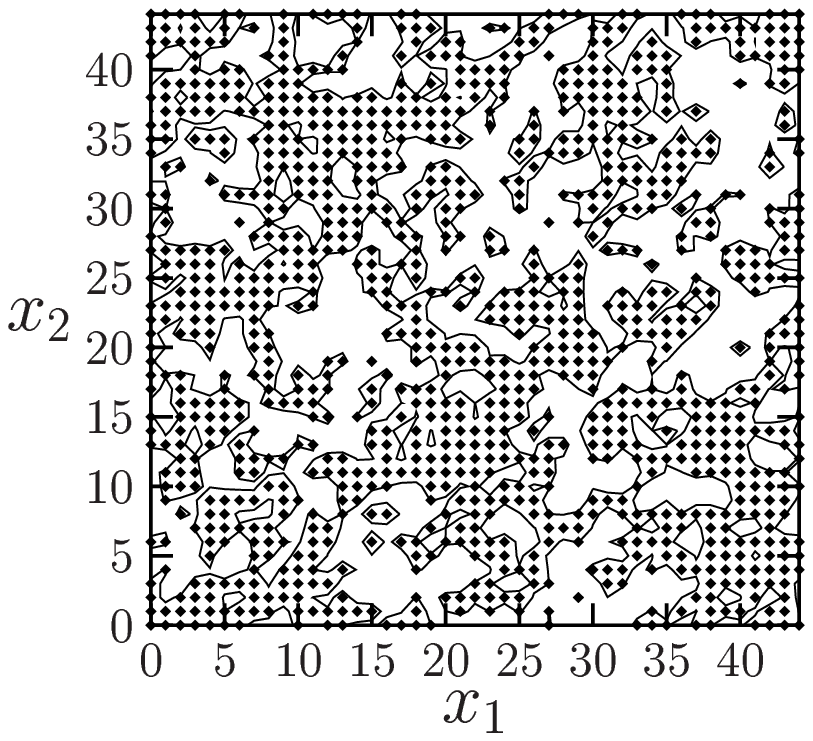}
\hspace{5mm}
\includegraphics[angle=0,width=0.33\linewidth]{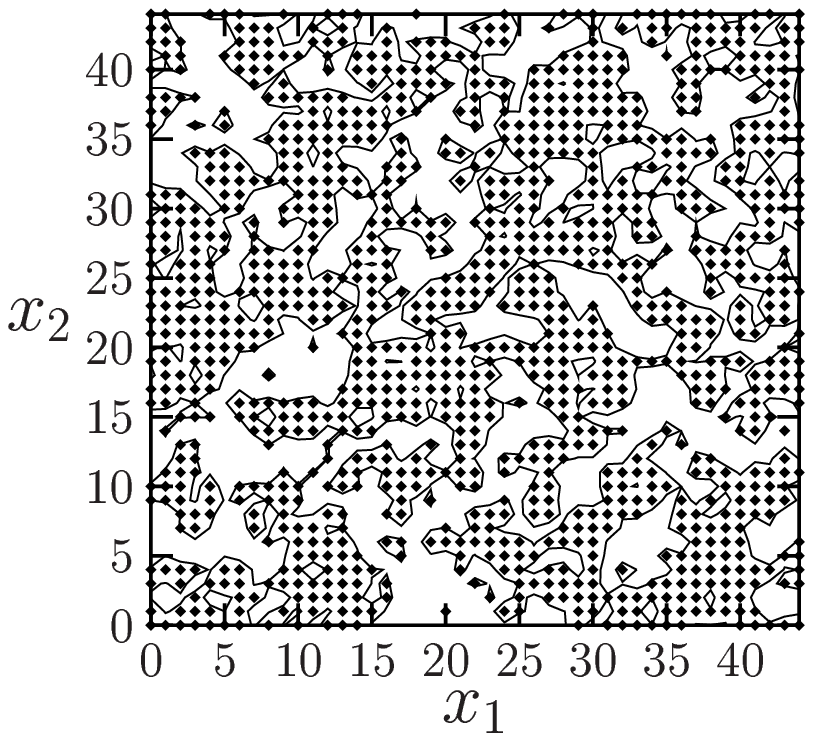} \\
\includegraphics[angle=0,width=0.33\linewidth]{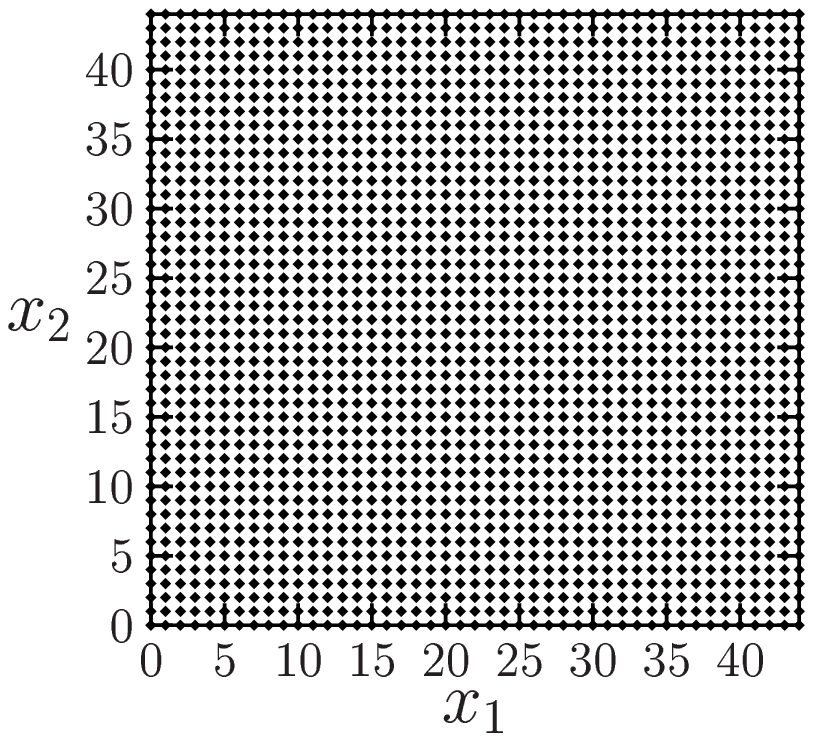}
\hspace{5mm}
\includegraphics[angle=0,width=0.33\linewidth]{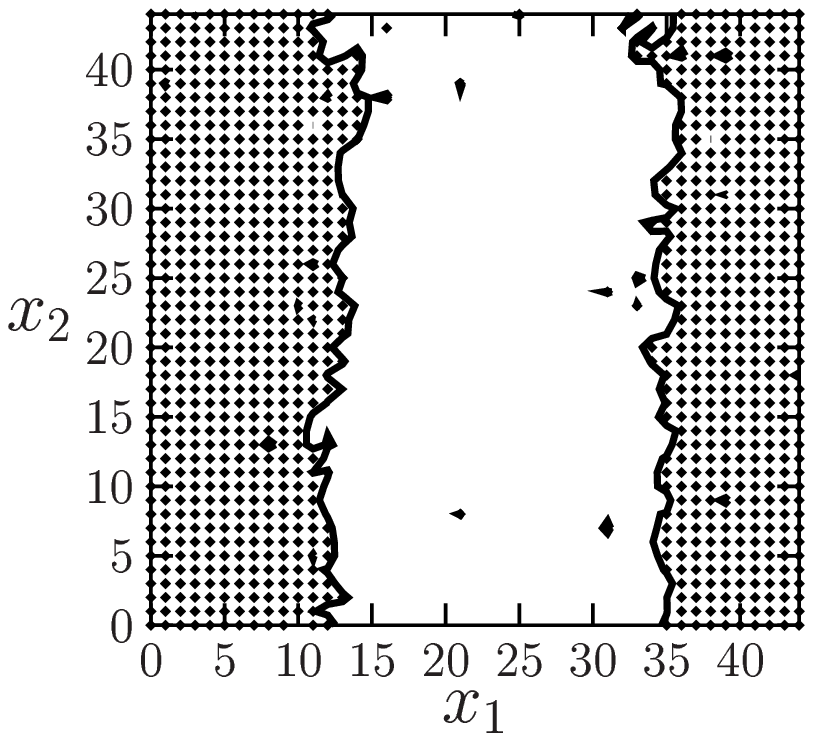}
\caption{Typical configuration of the 3d $\lambda \phi^{4}$ model
at $N=T=45$: we show the NC plane, and dark/bright areas correspond
to the two signs of $\phi$. The examples on the left (right) refer
to low (high) $\lambda$, and the upper (lower) plots were obtained
at weakly (strongly) negative $m^{2}$.}
\label{snap3d}
\end{figure}

A {\em quantitative} study was based on Monte Carlo simulations,
which probed the phase diagram in the $(m^{2},\lambda )$ plane,
for the 3d matrix formulation described in Subsection 2.1, at
$N = T = 15 \dots 45$ \cite{Frank}. Thus the picture by Gubser 
and Sondhi is converted into a uniform/disordered transition at 
small $\lambda$, and a striped/disordered transition at large
$\lambda$. This is in fact observed, as the phase diagram
in Figure \ref{phasedia3d} shows. Figure \ref{snap3d} adds
the features of typical configurations in the four sectors
(after mapping back the matrices to a lattice scalar field).
 
The phases, and their transitions, were identified with the
momentum dependent order parameter
\be \label{ordpar}
M(k) = \frac{1}{NT} \ \ ^{\rm ~~max}_{\frac{N}{2 \pi} \vert \vec p \vert =k} 
\ \Big\vert \sum_{t} \tilde \phi (\vec p , t) \Big\vert \ .
\ee
For $k=0$ this is the magnetisation, and for finite $k$
it captures the possible dominance of a ``stripe pattern'',
{\it e.g.}\ with $k$ parallel stripes, rotated in a suitable way
(if two non-zero components of $\vec p$ are involved, the
pattern is actually of a checker-board type).

Far from the transition lines, $M(k)$ indicates the phase unambiguously. 
The transition is identified best by varying $m^{2} < 0$ at fixed 
$\lambda$, and searching for a peak in the connected two-point function
\be
\la M(k)^{2} \ra_{\rm con} = \la M(k)^{2} \ra - \la M(k) \ra^{2} \ .
\ee
This provides accurate results for the critical values $m_{c}^{2}$.
For all $\lambda$ values that we considered, this transition
appears to be of second order. 

On the other hand, the transition uniform/striped inside the
ordered regime is rather hard to explore, as the uncertainty 
band in Figure \ref{phasedia3d} shows. Here we studied the
thermal cycle, which reveals a clear hysteresis effect \cite{Frank};
this is characteristic for a first order transition.\\

\begin{figure}[hbt!]
\center
\includegraphics[angle=0,width=0.5\linewidth]{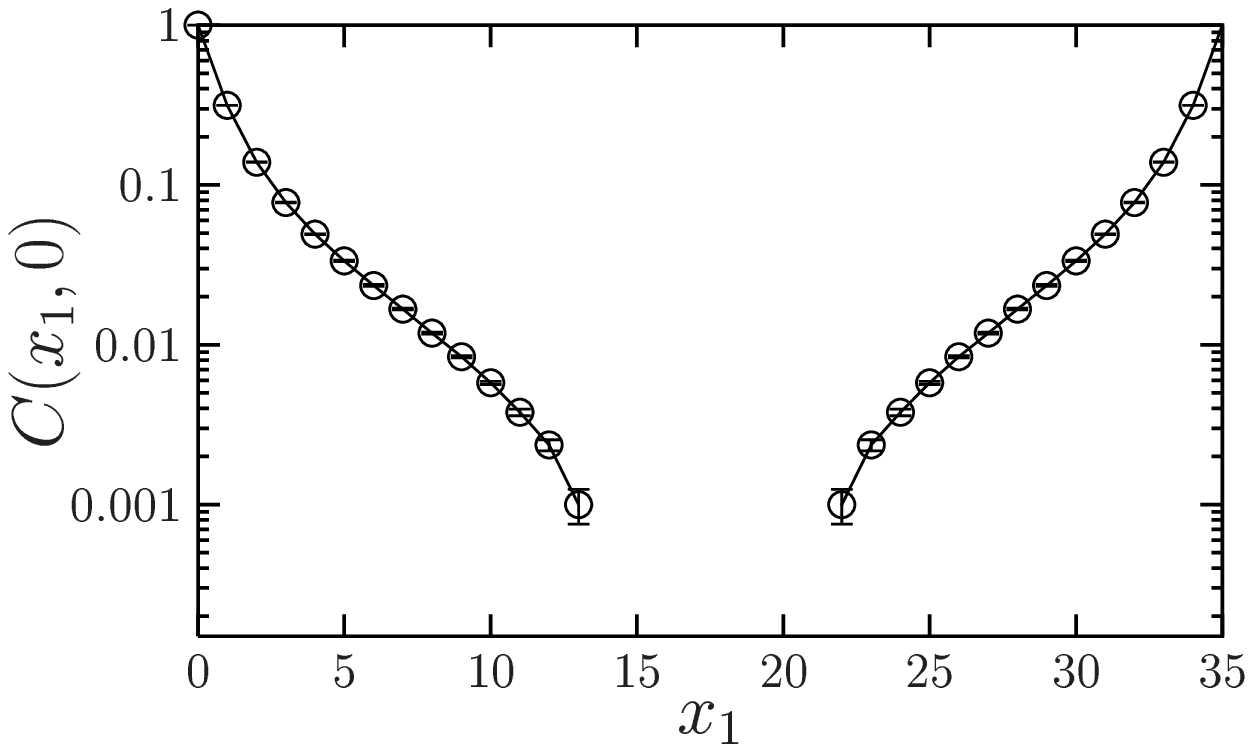}
\hspace{-3mm}
\includegraphics[angle=0,width=0.5\linewidth]{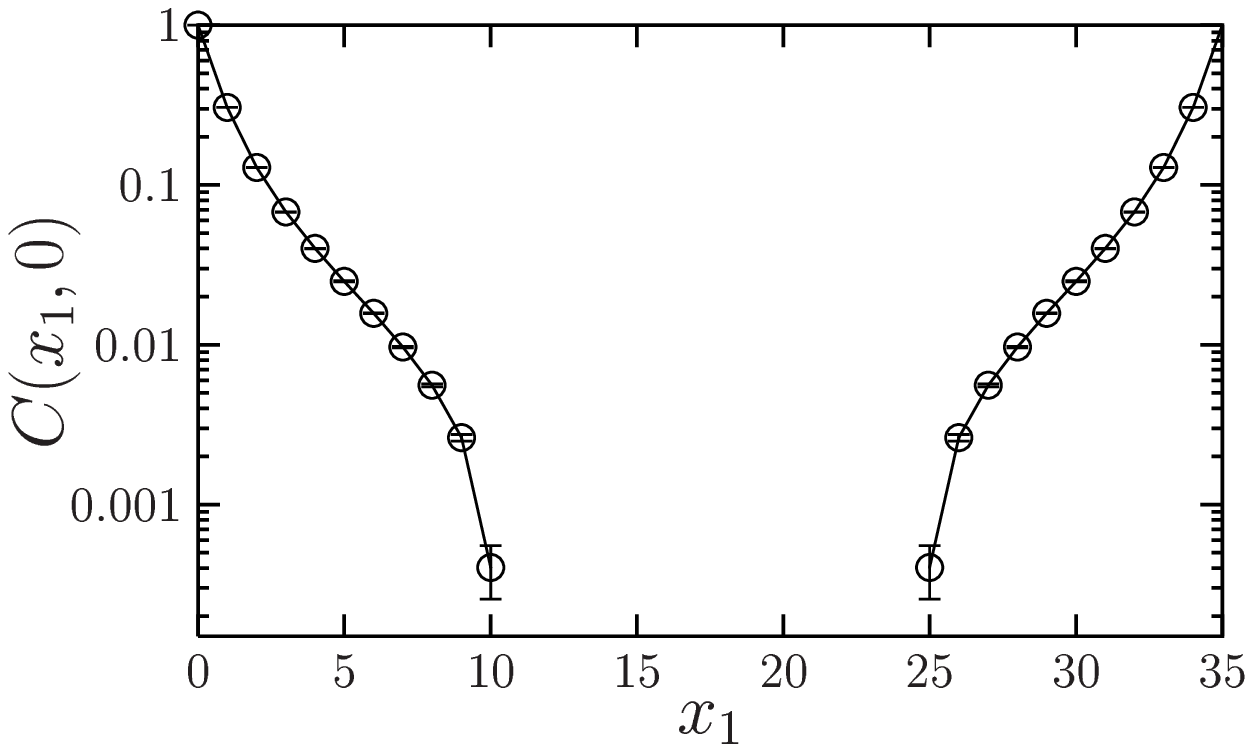}
\caption{The correlation function $C ( x_{1}, 0)$ for spatial 
separation, given in eq.\ (\ref{correspa}). Both plots were
obtained in the disordered phase, close to the ordering
transition, at $N=T=35$.
On the left the parameters are $N^{2}\lambda =70$, 
$N^{2}m^{2}=-17.5$ (close to the uniform phase), and on the right
$N^{2}\lambda =3500$, $N^{2}m^{2}=-140$ (close to the striped phase).
In both cases, the decay is fast, but not exponential.}
\label{spacorr}
\end{figure}

Next we considered the correlation functions close to the
order/disorder transition, in the disordered regime (where
finite size effects are harmless). 
Figure \ref{spacorr} refers to the spatial separation, and 
shows that the correlator
\be  \label{correspa}
C ( x_{1}, 0) = \la \phi (\vec 0 ; t) \, \phi (x_{1},0; t) \ra
\ee
has an unusual shape, both close to the uniform and close
to the striped phase. The decay is fast, but not exponential;
it is NC distorted.

\begin{figure}[hbt!]
\center
\includegraphics[angle=0,width=0.6\linewidth]{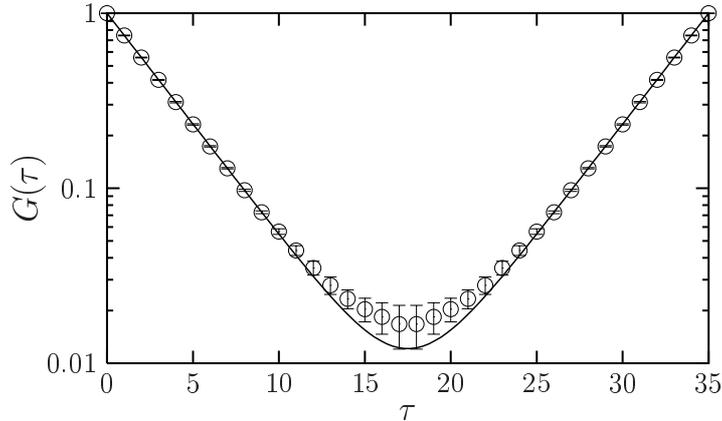}
\caption{The temporal correlation function $G( \tau )$, given in
eq.\ (\ref{corretime}), measured at $N=T=35$, $N^{2}\lambda = 350$,
$N^{2} m^{2} = -140$. The data agree very well with a cosh-fit.} 
\label{corretemp}
\end{figure}

Nevertheless we can evaluate the energy based on the {\em temporal} 
correlation function. Here we first transform the spatial
part of the configurations to momentum space, and we consider
\be  \label{corretime}
G( \tau ) = \la \tilde \phi (\vec p , t ) \, 
\tilde \phi (\vec p , t + \tau ) \ra \ .
\ee
Now we do find an exponential decay --- respectively a cosh
function behaviour at finite $T$ --- as Figure \ref{corretemp}
illustrates for the example $\vec p = \vec 0$.

\begin{figure}[hbt!]
\center
\includegraphics[angle=0,width=0.6\linewidth]{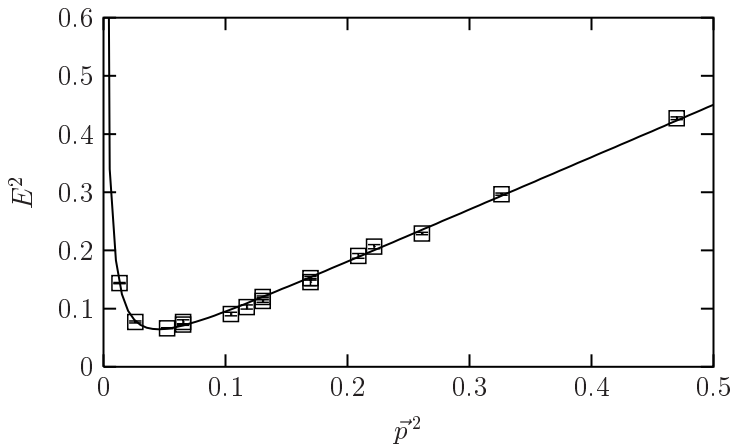}
\caption{The dispersion relation $E^{2}(\vec p^{\, 2})$
determined at $N = T = 55$, $m^{2} = -15$, $\lambda = 50$
(close to the striped phase). The energy takes its minimum
at finite momentum, which leads to a striped pattern at
somewhat lower $m^{2}$. A 4-parameter fit \cite{Frank}
is consistent with the trend to an IR divergence in 
infinite volume.}
\label{disprel}
\end{figure}

This leads to the dispersion relation $E^{2} (\vec p^{\, 2})$ 
shown in Figure \ref{disprel}, now for $N=T=55$. 
In fact, the energy minimum is located at {\em non-zero}
$\vec p$, which is a clear indication that we are close to
the striped phase; decreasing $m^{2}$ then leads to the
condensation of a corresponding striped pattern. It is evident 
that this pattern is non-uniform, but for the exact structure of 
the stripes various options are in close competition.

At very small momenta, the dispersion relation is consistent 
with the expected IR divergence.

\subsection{Double Scaling Limit}

So far we have been dealing with lattice units.
In order to take a continuum limit, we have to introduce
a scale, {\it i.e.}\ we need a dimensional reference quantity.
For this reason, we extrapolate
the (broad) linear regime in the dispersion
relation $E(| \vec p |)$ down to $\vec p = \vec 0$. This linear
extrapolation deviates from the dispersion at small momenta,
but it defines an effective mass $M_{\rm eff}$, according to
\be  \label{Meff}
E^{2} = M_{\rm eff}^{2} + \vec p^{\, 2} \ . 
\ee

\begin{figure}[hbt!]
\center
\includegraphics[angle=0,width=0.6\linewidth]{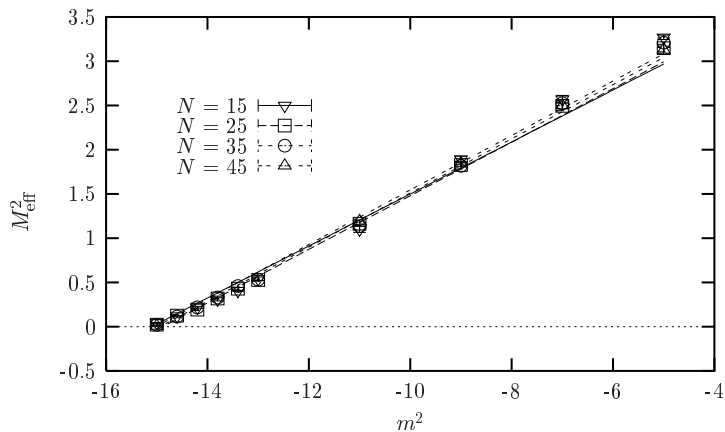}
\caption{The effective mass squared, $M_{\rm eff}^{2}$, obtained by
the extrapolation (\ref{Meff}) of the dispersion relation. We
see that $M_{\rm eff}^{2}$ depends linearly on $m^{2}$ resp.\
on $\Delta m^{2}$. The plot refers to $\lambda =50$, where
$M_{\rm eff}^{2}$ vanishes at $m_{c}^{2}=-15.01(8)$.}
\label{Meff2}
\end{figure}

We are going to investigate the behaviour if $m^{2}$ approaches
$m_{c}^{2}$ from above. Hence it is convenient to define
\be  \label{Dm2}
\Delta m^{2} := m^{2} - m_{c}^{2} \ .
\ee
At fixed $\lambda$, and for $\Delta m^{2} \gtrsim 0$,
we observed the proportionality relation 
\be
M_{\rm eff}^{2} \vert_{\lambda = {\rm const.}} \propto \Delta m^{2} \ ,
\ee
as Figure \ref{Meff2} shows for $\lambda =50$, as an example.

\begin{figure}[hbt!]
\center
\includegraphics[angle=0,width=0.6\linewidth]{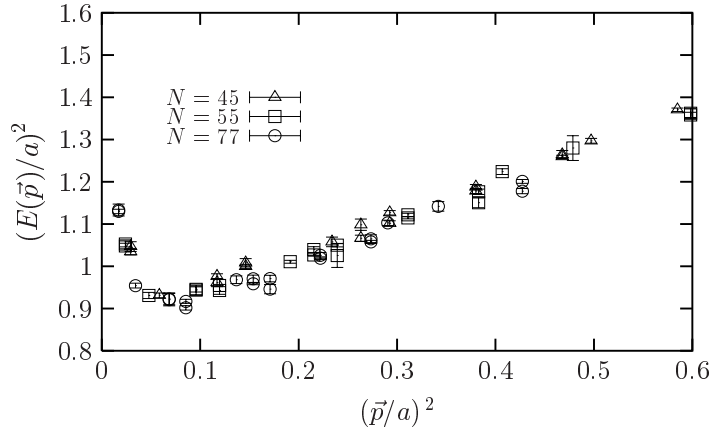}
\caption{Dispersion relation in dimensional units, at $\lambda = 50$ 
and $N \, \Delta m^{2} \simeq 100$. Both axes are given in units
of $M_{\rm eff}^{2}$. We see that the dispersion stabilises, 
so it {\em is} possible to take a DSL next to the striped phase, 
with a finite width of the stripes to be formed at $m^{2}< m_{c}^{2}\, $.}
\label{dispdim}
\end{figure}

Now we can take a continuum limit by keeping the effective
mass in dimensional units, $M_{\rm eff} / a$, fixed. For 
simplicity we set this ratio to $1$, hence $a= M_{\rm eff}^{-1}$ is
the dimensional lattice spacing. Therefore, the DSL condition
$\theta \propto Na^{2} = {\rm const.}$ is implemented by
increasing $N$, and simultaneously decreasing $m^{2}$ so
that it approaches $m_{c}^{2}$ from above, in such a manner that
\be
N \, \Delta m^{2} = {\rm const.} \ .
\ee
Now we can re-consider the dispersion relation in dimensional
units. The rest energy diverges like
$E(|\vec p | \to 0)/a \propto \sqrt{N}$ \ \cite{Frank},
which confirms the UV/IR mixing also dimensionally.
For a broad range of finite momenta $|\vec p |/a$, Figure 
\ref{dispdim} shows that the dispersion relation stabilises 
if we approach the DSL, and that the energy minimum is obtained 
around the dimensional momentum
\be
\vec p^{\, 2} / a^{2} \lesssim 0.1 ~ M_{\rm eff}^{2} \ .
\ee
This tells us that the striped phase persists in the DSL,
where stripes of finite width dominate.
This observation implies the spontaneous breaking of translation
and rotation symmetry in the striped phase of the 3d phase diagram. 

That phenomenon leads to a tricky question: does the
same happen also in $d=2$ ? The next section will be devoted
to this issue.

\section{Does translation symmetry break in $d=2$ ?}

The Mermin-Wagner Theorem \cite{MW} tells us that usually a continuous
global symmetry cannot break spontaneously in $d=2$. At first
sight this seems to imply that the striped phase cannot
occur in the NC plane. However, the proof for this theorem is
based on assumptions of standard quantum field theory,
like locality --- this does not hold on the NC plane.

Still, Gubser and Sondhi did not expect a striped phase
in $d=2$ \cite{GuSo}. They presented a consideration how 
the Mermin-Wagner Theorem could be extended even to NC field
theory. They used an effective action approach of the
Brazovskii-type, where the kinetic term is of quartic
order in the momentum, which should make the exclusion
of spontaneous symmetry breaking stronger.
On the other hand, the effective action approach of Ref.\
\cite{CaZa08} seems to affirm a striped phase.

From the numerical side, a striped phase in the NC
$\lambda \phi^{4}$ model has been manifestly observed also
in $d=2$ \cite{proc,aristocat,Frank}. However, this does not
prove its existence in the DSL --- it could also be
an artifact of the lattice and of finite volume.
The fate of this phase in the DSL has been investigated
numerically only very recently \cite{Hector}.

\begin{figure}[hbt!]
\center
\includegraphics[angle=0,width=0.8\linewidth]{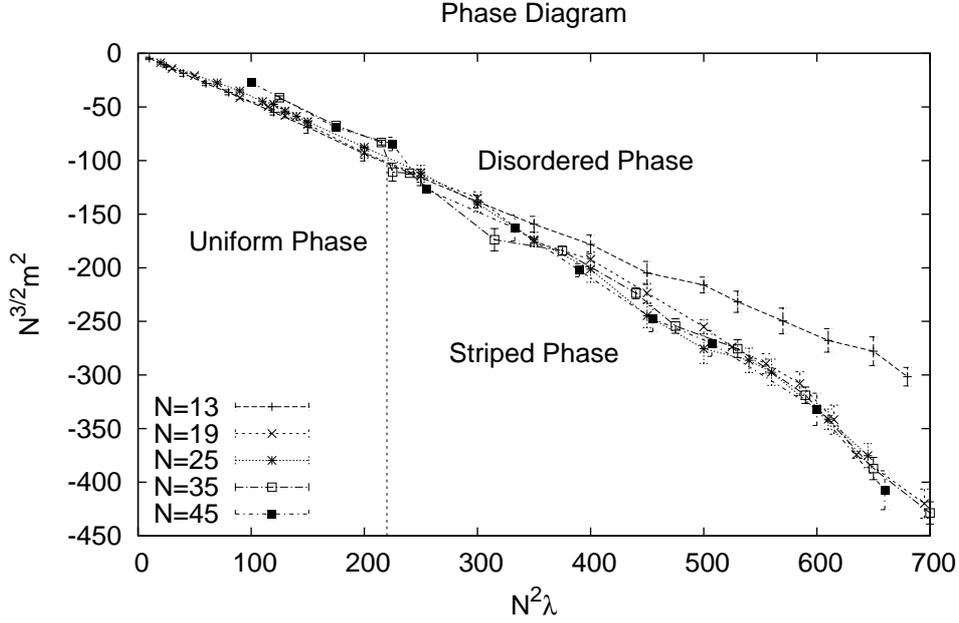}
\caption{The phase diagram of the 2d NC $\lambda \phi^{4}$ model.
In contrast to the 3d case, the vertical axis has to be chosen 
as $N^{3/2} m^{2}$. Then the transition line between disorder
and the ordered phases stabilises at $N \geq 19$.}
\label{phasedia2d}
\end{figure}

The matrix model formulation corresponds to the description
in Subsection 2.1, where the time direction collapses to one 
site. Figure \ref{phasedia2d} shows the phase diagram in $d=2$.
We see that the vertical axis needs a scaling factor $N^{3/2}$, 
which differs from the 3d case (cf.\ Figure \ref{phasedia3d}). 
If $m^{2}$ is scaled in this way, we observe a convincing stabilisation
of the order/disorder transition line for $N \geq 19$.

\begin{figure}[hbt!]
\center
\includegraphics[angle=0,width=0.45\linewidth]{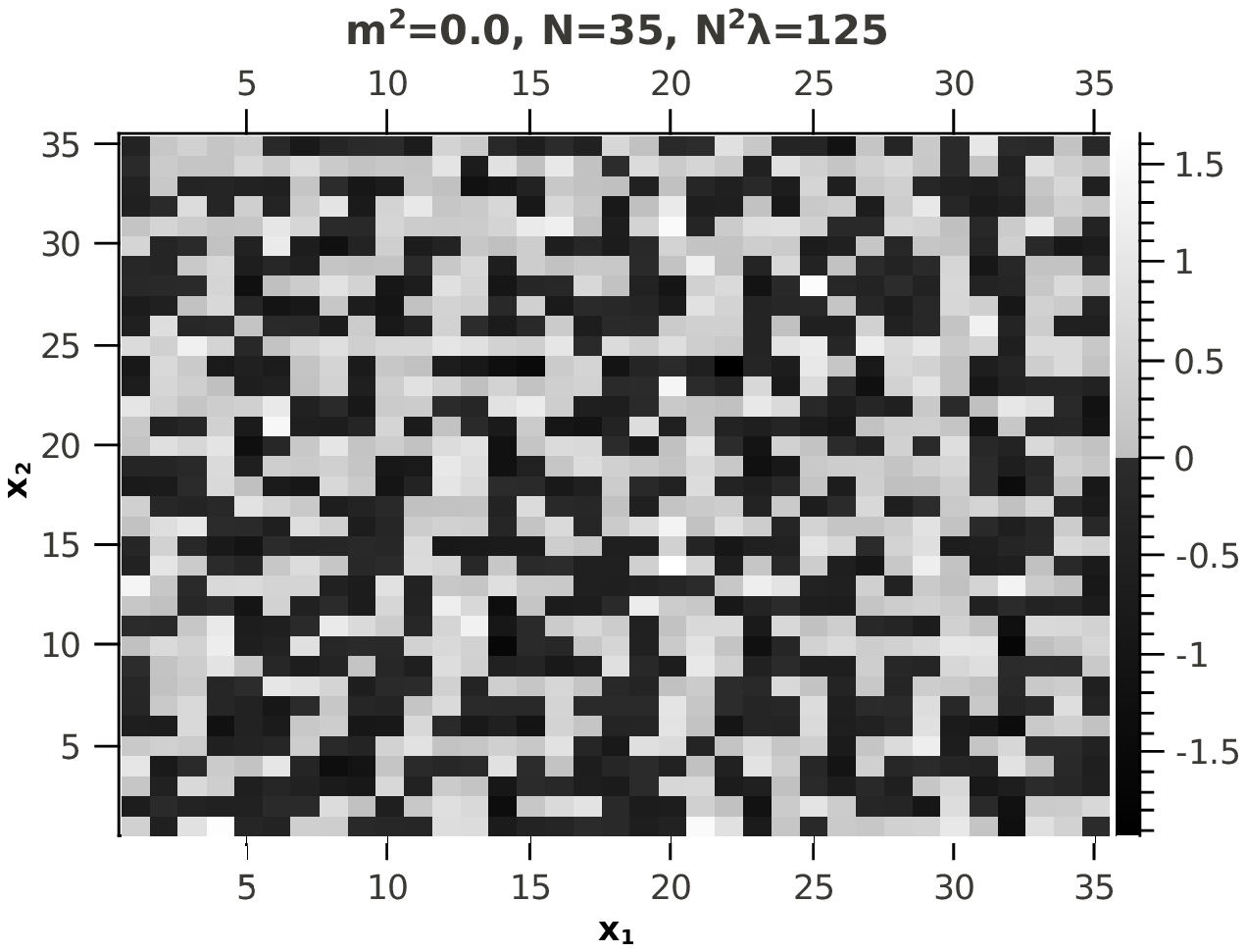}
\hspace{-3mm}
\includegraphics[angle=0,width=0.45\linewidth]{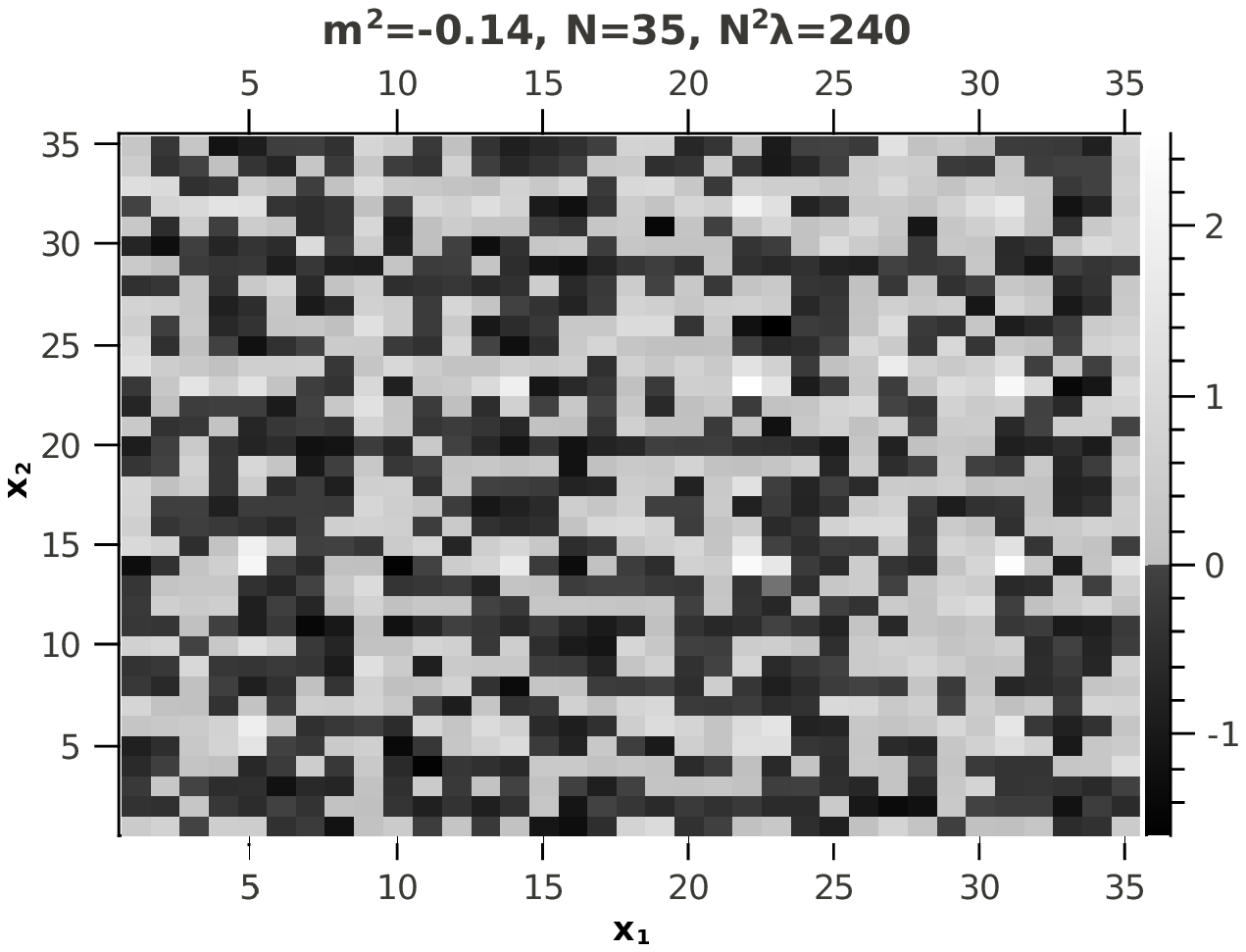}
\includegraphics[angle=0,width=0.45\linewidth]{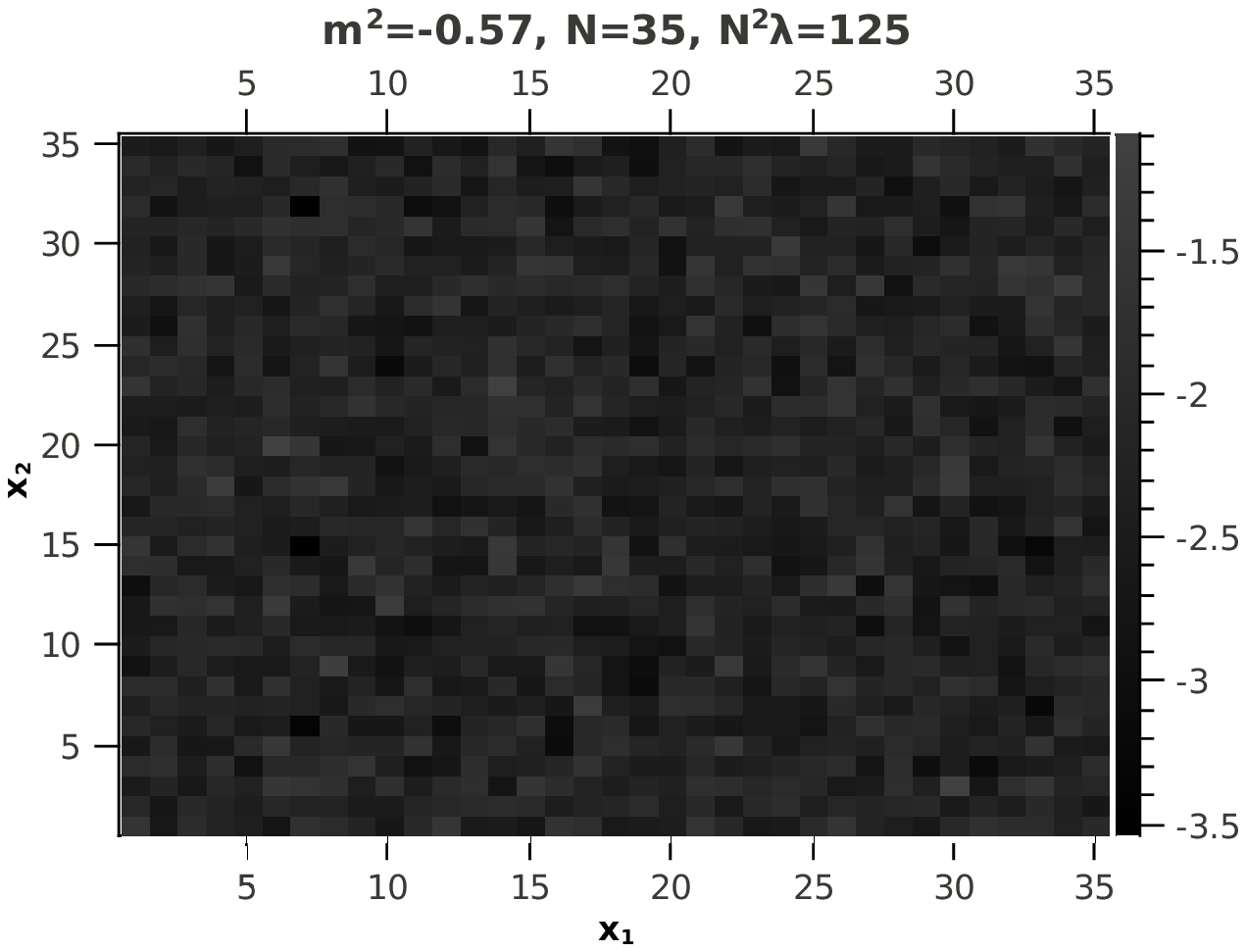}
\hspace{-3mm}
\includegraphics[angle=0,width=0.45\linewidth]{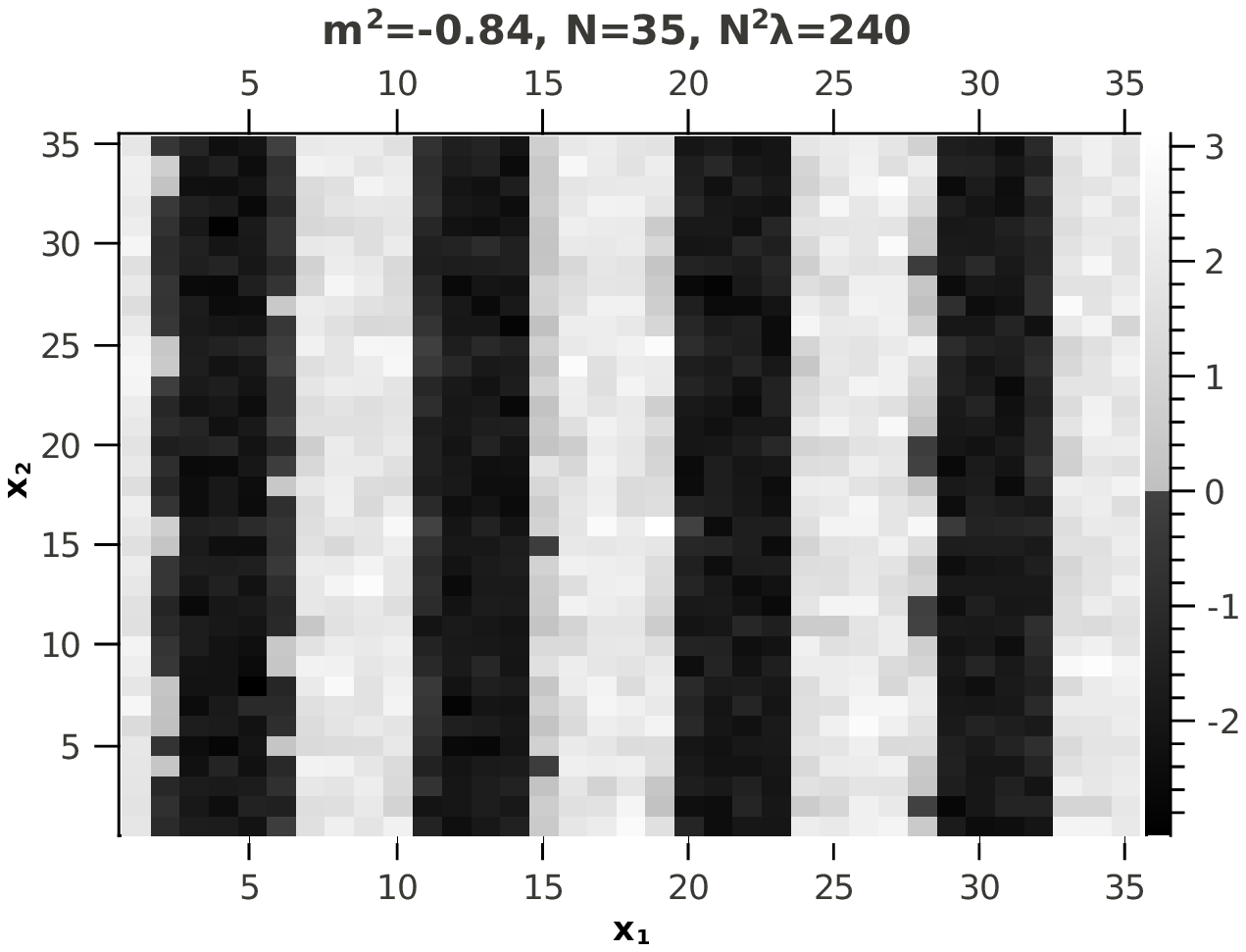}
\caption{Typical configurations in four sectors of the phase
diagram in Figure \ref{phasedia2d}, analogous to Figure \ref{snap3d}.
The upper plots are in the disordered phase, next to the uniform
and to the striped phase, respectively. The lower plots are examples
for uniform and for striped ordering.}
\label{snap2d}
\end{figure}
Figure \ref{snap2d} shows again typical configurations in
the four sectors of this phase diagram, in analogy to the
Figure \ref{snap3d}, but now with a four-stripe pattern.
\begin{figure}[hbt!]
\center
\includegraphics[angle=0,width=0.47\linewidth]{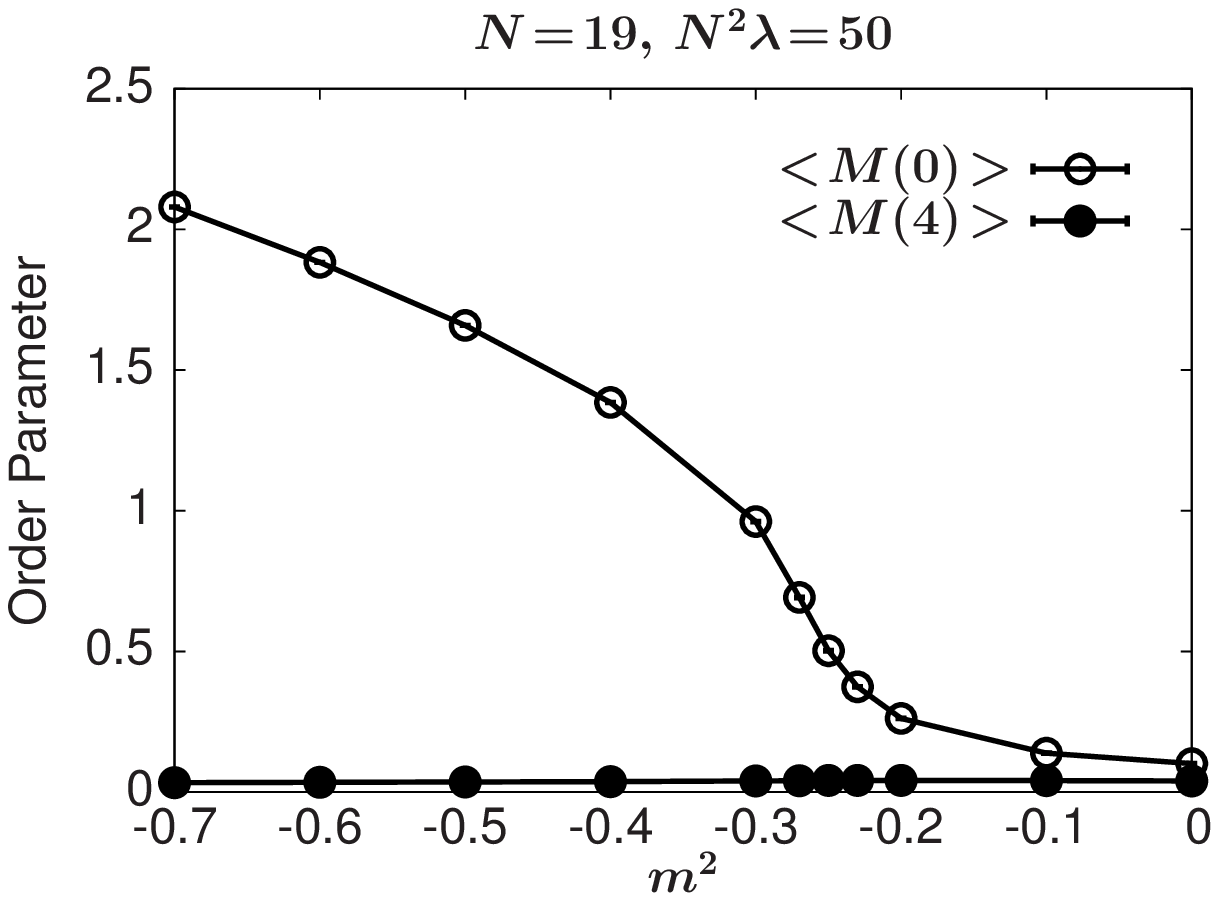}
\hspace{-3mm}
\includegraphics[angle=0,width=0.47\linewidth]{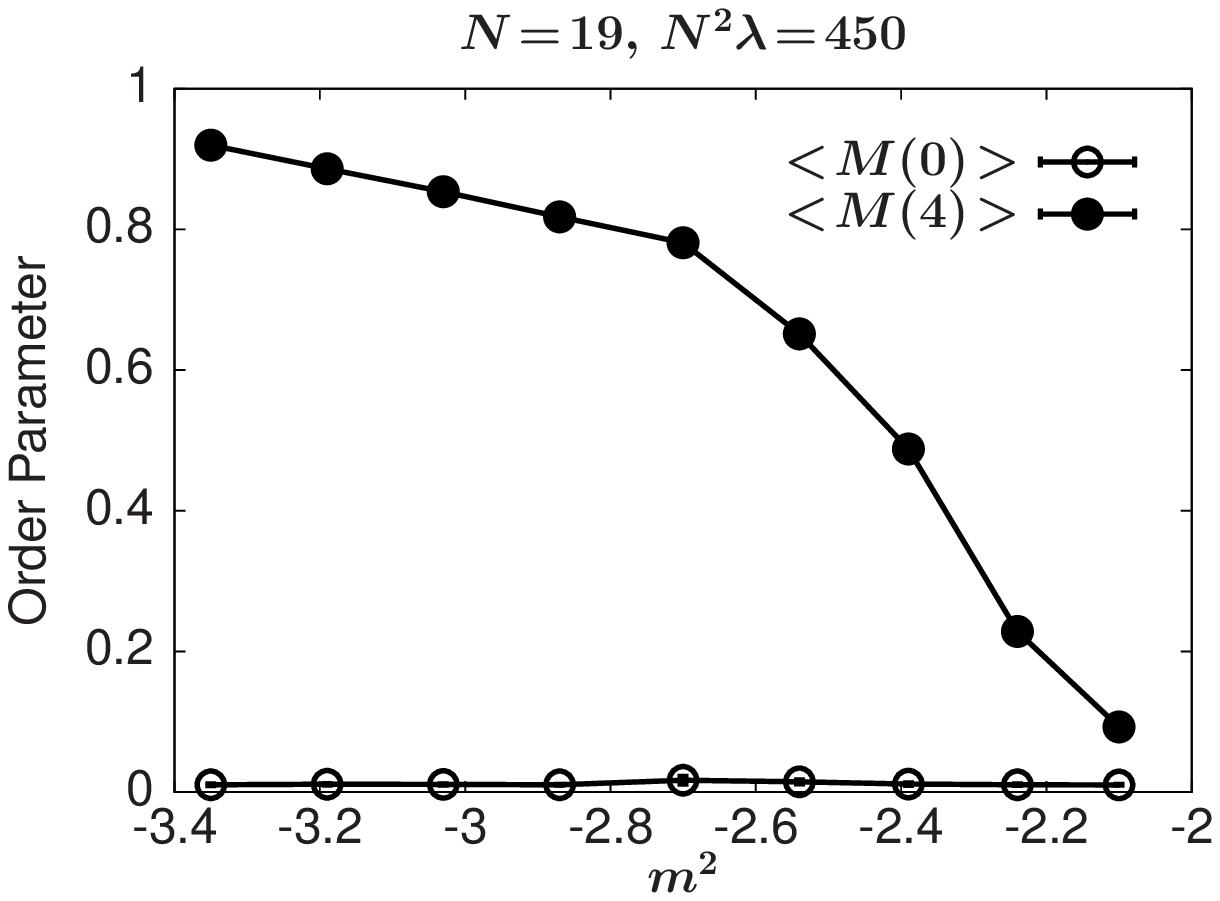} \vspace*{2mm} \\
\includegraphics[angle=0,width=0.47\linewidth]{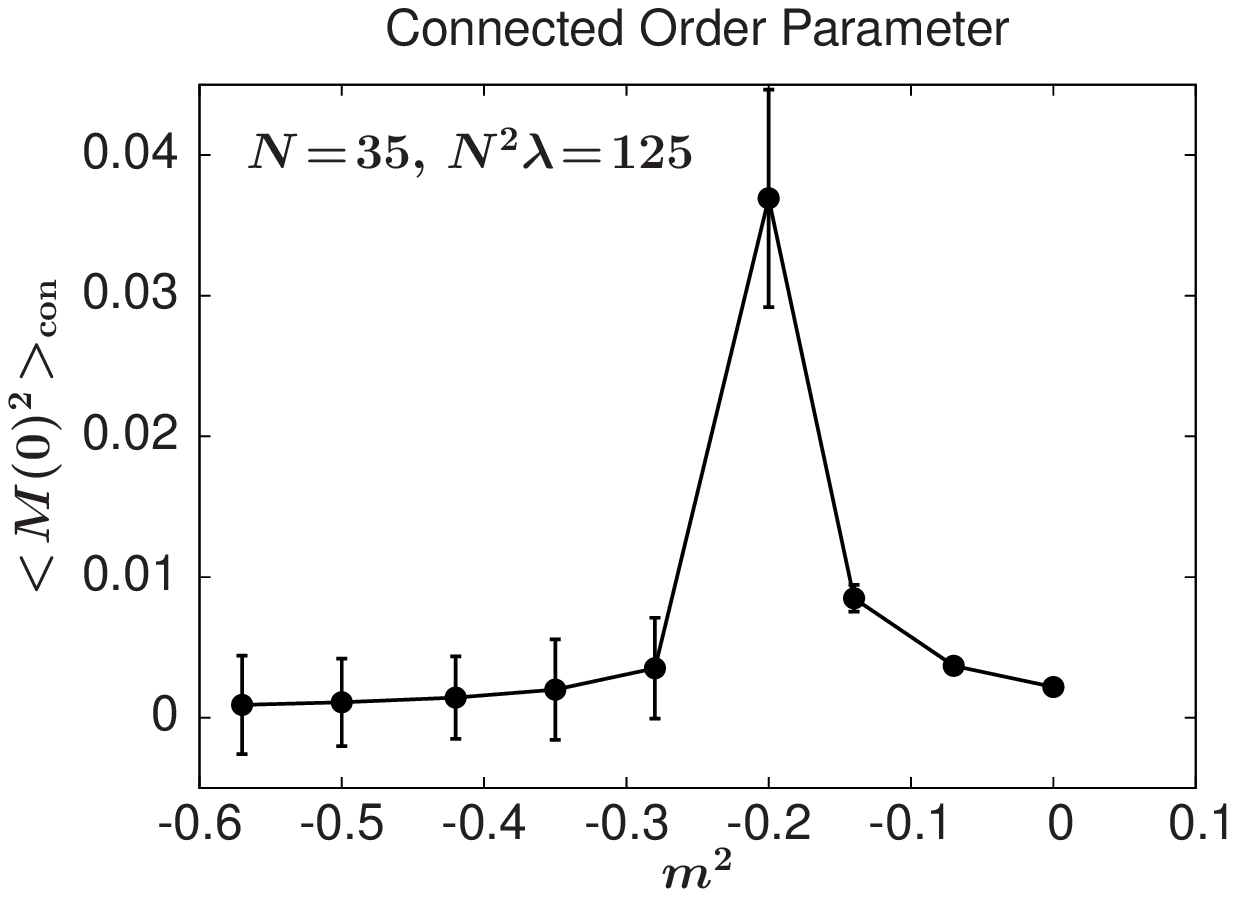}
\hspace{-3mm}
\includegraphics[angle=0,width=0.47\linewidth]{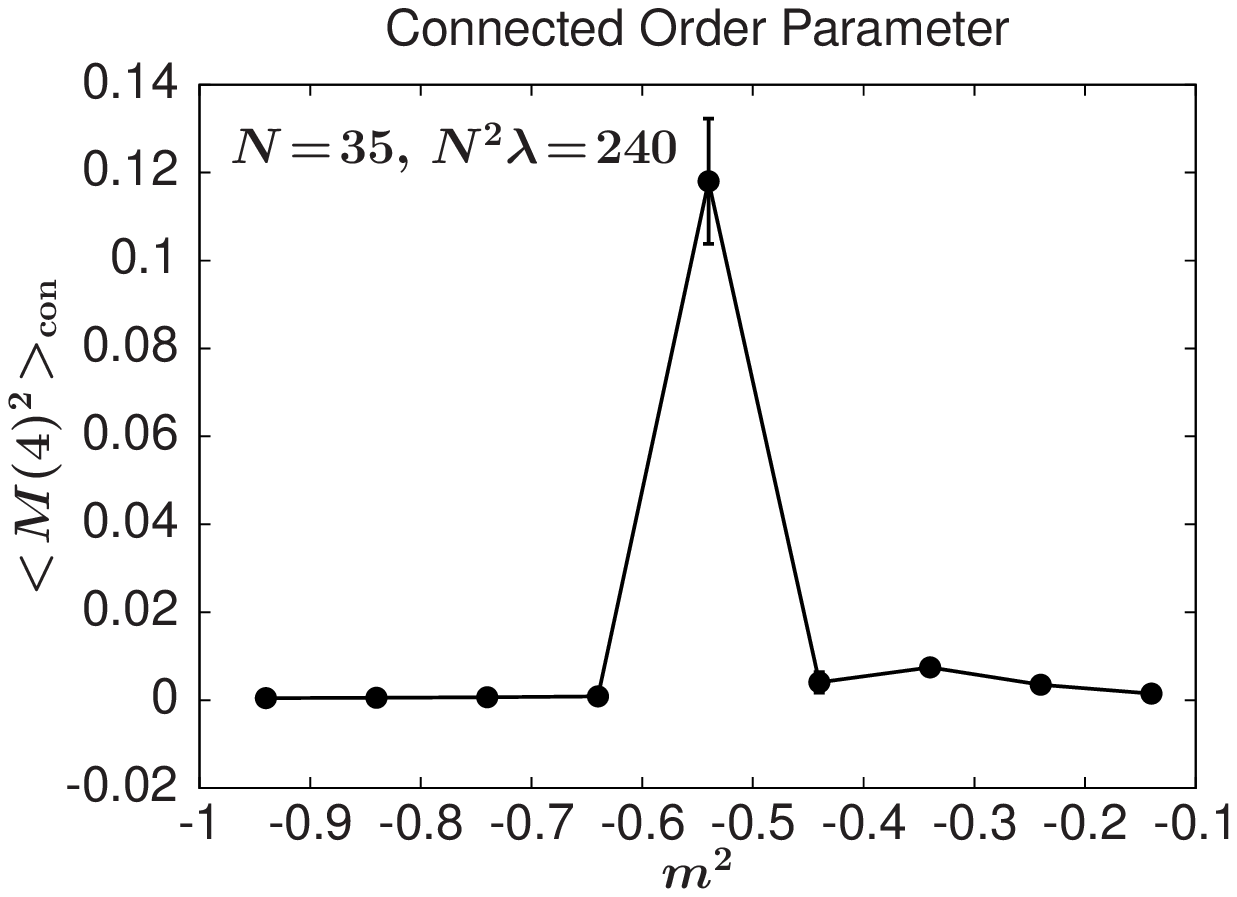}
\caption{The order parameters $\la M(0)\ra$ and $\la M(4) \ra$, as 
defined in eq.\ (\ref{ordpar}), for the 2d NC $\lambda \phi^{4}$ model. 
We keep $N$ and $\lambda$ fixed and show the dependence on $m^{2}$. 
Above we see that for small (large) $\lambda$ and decreasing $m^{2}$
the disorder turns into a uniform (4-stripe) pattern. Below we
show examples where the 2-point functions of these order parameters
have a peak, which allows us to identify the critical value $m^{2}_{c}$.}
\label{ordpara}
\end{figure}
The transition was again detected with the order parameter
(\ref{ordpar}). Figure \ref{ordpara} gives examples how the 
uniform or striped order parameter rises for decreasing $m^{2}$,
and how the corresponding connected correlator exhibits
a peak at the transition.

\begin{figure}[hbt!]
\center
\includegraphics[angle=0,width=0.55\linewidth]{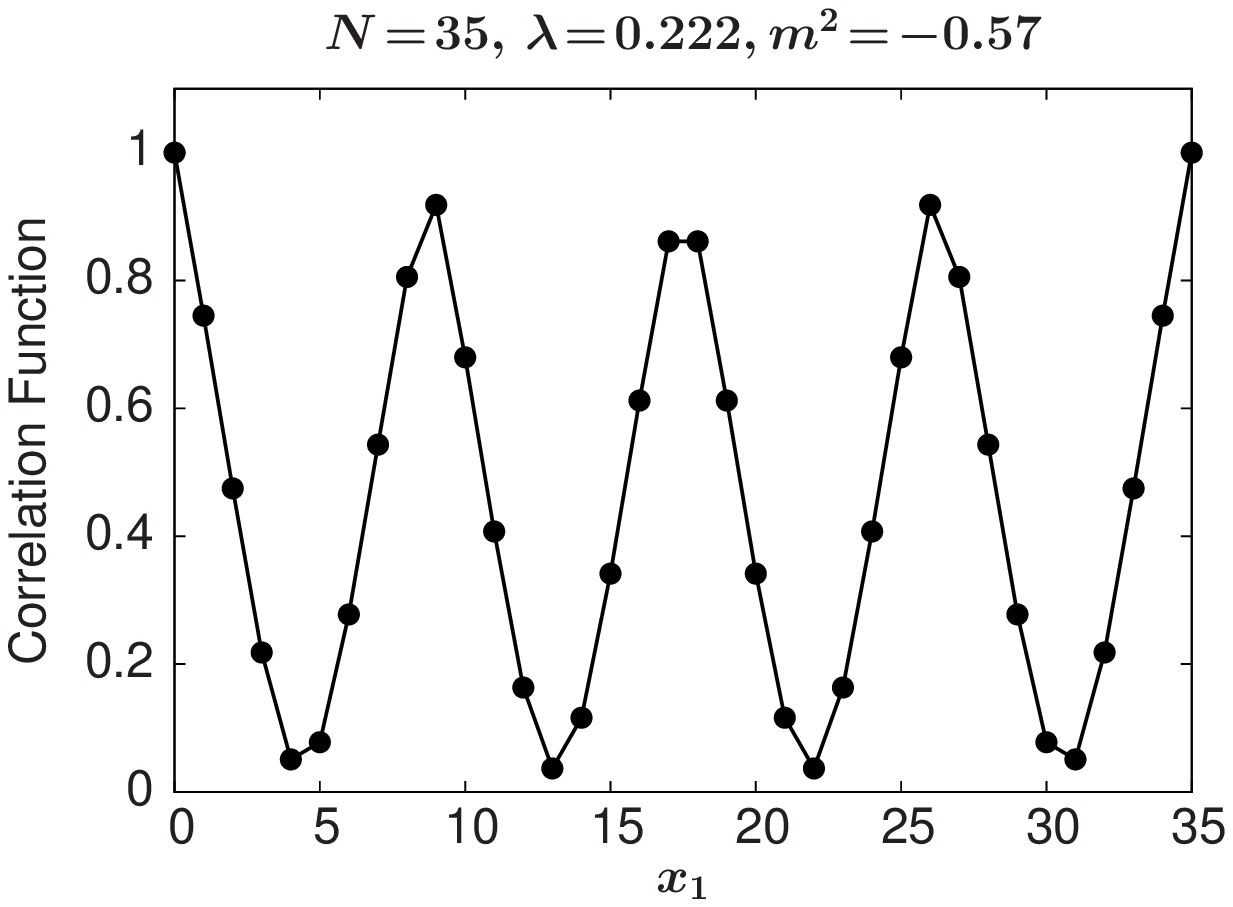}
\caption{The correlation function $\la \phi_{(0,0)} \phi_{(x_{1},0)}\ra$
near the striped phase, at $(N^{3/2}m^{2},N^{2} \lambda ) = (-118,272)$.
For somewhat lower $m^{2}$ a 4-stripe pattern condenses.}
\label{corre2d}
\end{figure}
Now we consider also here the correlation function. In $d=3$
we focused on the correlation in time direction in order to
extract the dispersion relation and to introduce the effective
mass $M_{\rm eff}$, as a scale for the DSL. This is not available
anymore in $d=2$, so now we have to deal with the spatial
correlation, and its unusual decay behaviour. Figure \ref{corre2d}
shows an example in the disordered phase, but next to the
the striped phase --- for slightly lower $m^{2}$ a 4-stripe
pattern will condense (like the example in Figure \ref{snap2d}
on the bottom at the right).

Our concept is as follows: we decrease $m^{2}$ down towards
$m^{2}_{c}$, and we increase the matrix size $N$ at the same time,
such that the decay of the correlator stabilises down to the first
dip. This replaces the usual reference to the exponential decay.
The difference $\Delta m^{2}$, defined in eq.\ (\ref{Dm2}), 
introduces a scale, which translates --- with a suitable exponent 
--- into the scale of the DSL, 
\be
a^{2} \propto ( \Delta m^{2} )^{\sigma} \ .
\ee
The exponent $\sigma$ has to be identified, then we can
address the question whether or not it is possible to take
a DSL and keep close to the striped phase. If this
can (cannot) be done, this is evidence for the existence 
(absence) of the striped phase in the continuum and
infinite volume limit of this model.

To tackle this question, we first choose a normalisation by setting
the lattice spacing for $N=35$ to $a=1$. Thus the DSL converts
a lattice distance $x$ into the dimensional distance
\be  \label{dimdist}
a x = \sqrt{\frac{35}{N}} \, x \ .
\ee
This distance should be compatible up to the first correlation
dip for increasing $N$.

In analogy to the ``dimensionless temperature'', which is often
used near a phase transition, $(T - T_{c})/T_{c}$, we adjust the
dimension by the suitable power of $m_{c}^{2}$, such that the
DSL can be written as
\be
N a^{2} = N \frac{(\Delta m^{2})^{\sigma}}{(m_{c}^{2})^{1 - \sigma}} \ .
\ee

\begin{figure}[hbt!]
\center
\includegraphics[angle=0,width=0.5\linewidth]{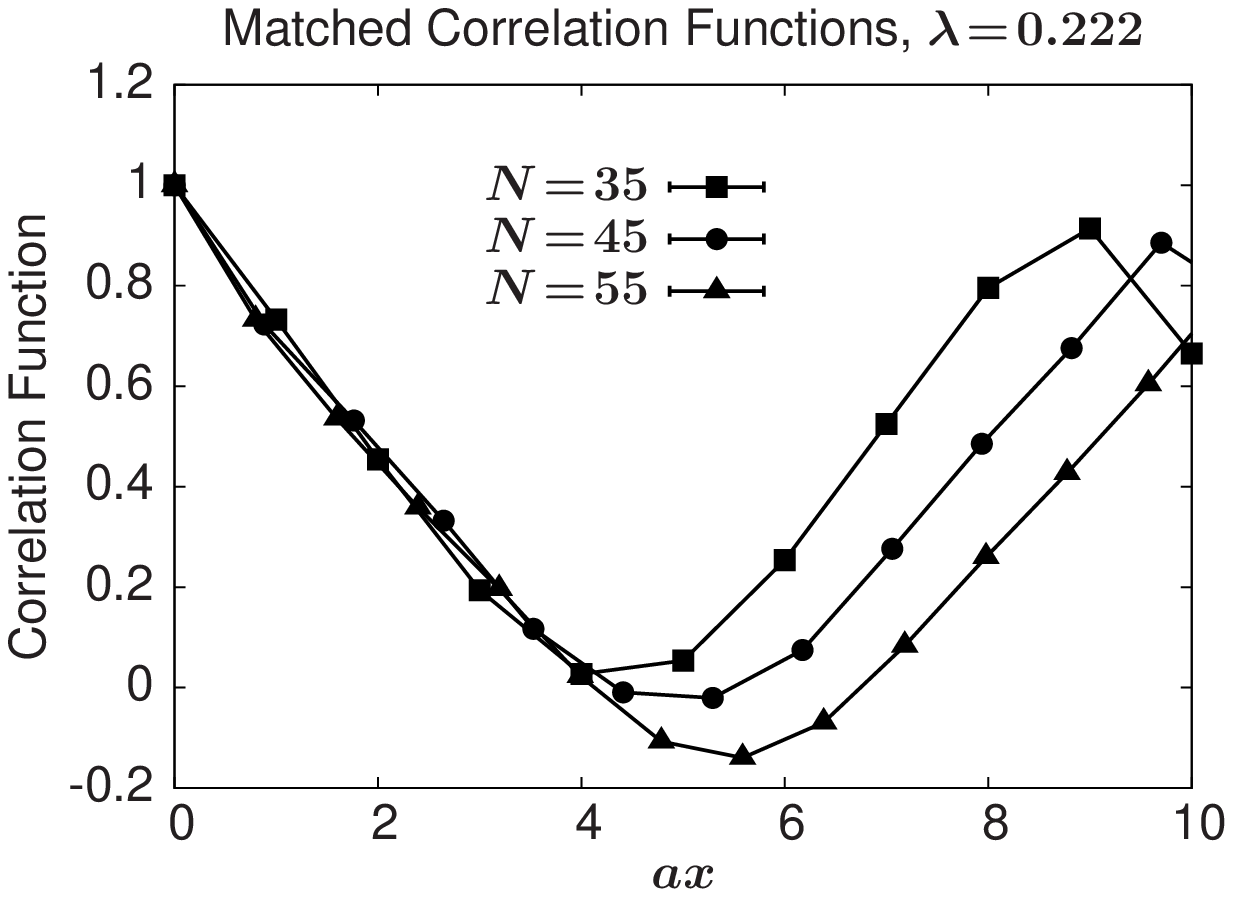} \vspace*{3mm} \\
\includegraphics[angle=0,width=0.5\linewidth]{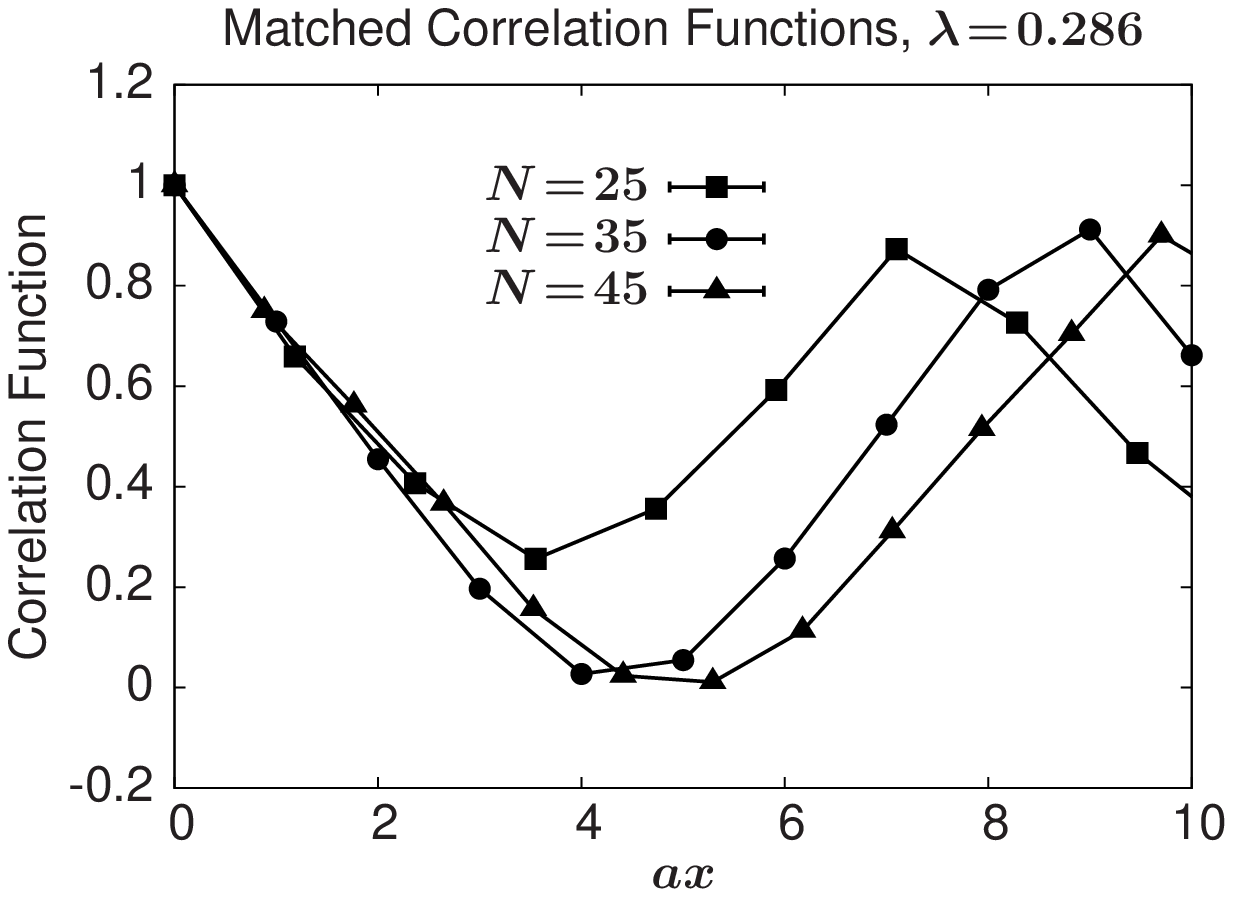} \vspace*{3mm} \\
\includegraphics[angle=0,width=0.5\linewidth]{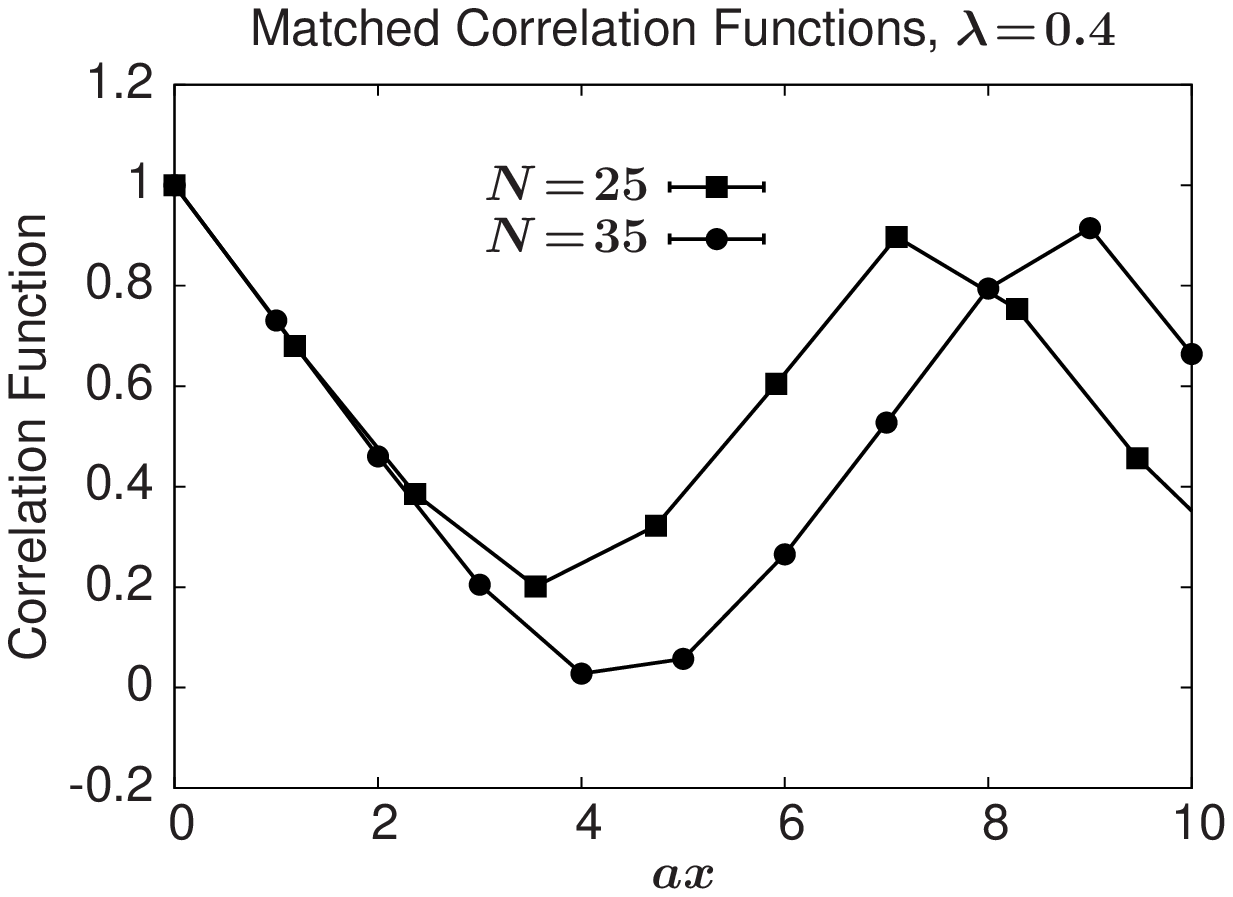}
\caption{Three examples for ``matched correlation functions'': at
fixed $\lambda$, but for different sizes $N$, the parameter 
$\Delta m^{2} = m^{2} - m^{2}_{c}$ is tuned such that the correlation
decay down to the first dip has the same slope. Then the distance in 
physical units --- as given in eq.\ (\ref{dimdist}) --- agrees. In this 
way we identify $\Delta m^{2}$ values to be inserted in eq.\ (\ref{sig}),
which determines the exponent $\sigma$.}
\label{corrematch}
\end{figure}
Now let us consider two matrix sizes $N_{1}$ and $N_{2}$; we want
to identify the mass shifts $\Delta m^{2}_{1}$ and $\Delta m^{2}_{2}$
which correspond to the same trajectory towards the DSL.
We fix $\lambda$ to the same value, so that the dimensionless
term $\lambda \theta$ remains constant. We fine-tune the mass
shifts such that they lead to the same short-distance correlation
decay. This is illustrated for three examples in Figure 
\ref{corrematch}. Once these values $\Delta m^{2}_{i}\,$ $(i=1,2)$,
and the corresponding critical values $m^{2}_{c,i}$, are determined,
we extract the exponent $\sigma$ from
\be  \label{sig}
\sigma = \frac{\ln (m_{\rm 1,c}^{2} / m_{\rm 2,c}^{2})}
{\ln (\Delta m_{\rm 1,c}^{2} / \Delta m_{\rm 2,c}^{2})
+ \ln (m_{\rm 1,c}^{2} / m_{\rm 2,c}^{2})} \ .
\ee
This has to be done for a variety of pairs $N_{1}$, $N_{2}$, and the 
crucial question is whether or not a stable $\sigma$-value is obtained.
\begin{table}[h!]
\centering
\begin{tabular}{|c|c|c||c|}
\hline
$\lambda$ & $N_{1}$ & $N_{2}$ & $\sigma$ \\ 
\hline
\hline
\multirow{3}{*}{0.222} & 35 & 45 & 0.152(7) \\
  & 35 & 55 & 0.156(6) \\
  & 45 & 55 & ~ 0.161(11) \\
\hline
\multirow{3}{*}{0.286} & 25 & 35 & 0.161(9) \\
 & 25 & 45 & 0.167(7) \\
 & 35 & 45 & ~ 0.178(23) \\
\hline
 0.4 & 25 & 35 & ~ 0.147(13) \\
\hline
\end{tabular}
\caption{The $\sigma$-values obtained for various pairs of
sizes $N_{1}$, $N_{2}$ after tuning $\Delta m^{2}$ such that the
short-distance decay of the correlation function coincides.}
\label{tabsigma}
\end{table}

\begin{figure}[hbt!]
\centering
\includegraphics[angle=0,width=0.6\linewidth]{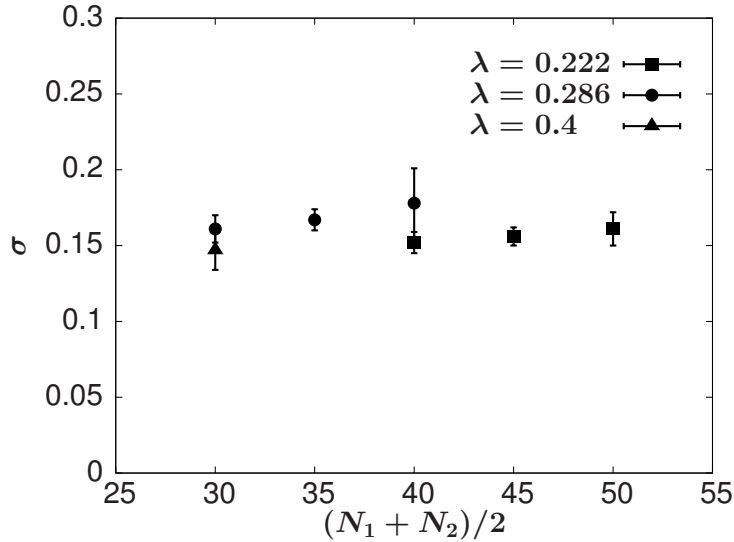}
\caption{An illustration of the values given in Table \ref{tabsigma}.
We see a clear trend to a plateau value of $\sigma = 0.16(1)$.}
\label{sigmaplot}
\vspace*{-2mm}
\end{figure}
There are practical constraints for these evaluations:
$\lambda$ has to be large enough to be close to the
striped (not uniform) phase for the smaller $N$ involved.
On the other hand, for the larger $N$ the product $N^{2} \lambda$ 
should not be too large --- otherwise we run to the far right 
in the phase diagram in Figure \ref{phasedia2d},
where the effective potential forms a landscape of many
deep valleys (meta-stable local minima), so reliable simulations 
are more and more difficult to achieve (the trouble starts already 
with the thermalisation).

Still a numerically accessible window could be found, and we
give our results for three $\lambda$ values in Table \ref{tabsigma}
and Figure \ref{sigmaplot}.
We considered the uncertainties which affect $\sigma$ 
(errors in $\Delta m^{2}_{i}$ and in $m^{2}_{c,i}$), but the precision
is sufficient to confirm a clear trend towards a stable exponent
\be
\sigma = 0.16(1) \ .
\ee
Thus we can indeed approach the DSL consistently, running to the 
right in the phase diagram in Figure \ref{phasedia2d},
while staying in the vicinity of the striped phase.
This implies that the latter does persist in the DSL.
Hence in the NC world translation symmetry can indeed
break spontaneously, even in $d=2$.

\section{Conclusions}

We have studied the $\lambda \phi^{4}$ model in $d=3$ \cite{Frank}
and in $d=2$ \cite{Hector}. In both cases, the spaces include a 
non-commutative plane. In order to explore this model beyond 
perturbation theory, we introduced a (fuzzy) lattice regularisation
and mapped the theory onto a Hermitian matrix model,
following Ref.\ \cite{AMNS}. This enables Monte Carlo simulations,
which were performed with a Metropolis algorithm.\footnote{There
have been related studies of the $\lambda \phi^{4}$ model on a {\em 
fuzzy sphere} instead of a NC plane \cite{fuzzysphere}. Also in that case 
Monte Carlo simulations were performed after the mapping onto 
a Hermitian matrix model. However, the non-commutativity relation 
differs from the form (\ref{NCplane}) that we have discussed here.}

For both dimensions we observed that a strongly negative
bare mass parameter $m^{2}$ enforces some order. 

\begin{itemize}

\item If $\lambda$ is small, so that non-commutativity effects 
are weak, this order is the uniform magnetisation, as in the commutative
variant of this model.

\item If $\lambda$ is large, so that non-commutativity effects 
are strongly amplified, this order is ``striped''. That phase
does not occur in the commutative case.

\end{itemize}

In both dimensions we gave numerical evidence
that the striped phase persists in the Double Scaling Limit,
which extrapolates simultaneously to the continuum and
to infinite volume, while keeping the non-commutativity
parameter $\theta$ fixed. This indicates the spontaneous
breaking of translation and rotation symmetry in this limit.

For the 2d case this might appear surprising due to the
Mermin-Wagner Theorem. However, this theorem does not
apply to non-commutative field theory, since it assumes
locality and an IR regular behaviour. \\

\ack Part of the results presented here are based on our
collaboration with Jun Nishimura. We thank him, as well as 
Antoniox Bigarini and Jan Volkholz, for their contributions,
and Urs Gerber for reading the manuscript.

This work was supported by the Mexican {\it Consejo Nacional de Ciencia 
y Tecnolog\'{\i}a} (CONACyT) through project 155905/10 ``F\'{\i}sica 
de Part\'{\i}culas por medio de Simulaciones Num\'{e}ricas'',
and by the Spanish {\it MINECO} (grant SEV-2012-0249).
The recent simulations were performed on the cluster of the 
Instituto de Ciencias Nucleares, UNAM. \\ \ \\

\end{document}